\documentclass{aa} 
\bibpunct{(}{)}{;}{a}{}{,}
\usepackage[squaren,Gray]{SIunits}
\usepackage{color}
\usepackage{lmodern}
\usepackage{graphicx}
\usepackage{txfonts}
\usepackage{natbib}


\definecolor{forestgreen}{rgb}{0.11,0.54,0.15}
\definecolor{purple}{rgb}{0.62,0.10,0.96}

\def\pfactor{1.26}
\def\pfactorerrstat{0.04}
\def\pfactorerrsyst{0.06}
\def\pfactorerrtot{0.07}

\begin{document} 

   \title{Observational calibration of the projection factor of Cepheids}

   \subtitle{I. The Type II Cepheid $\kappa$\,Pavonis\thanks{Based on observations realized with ESO facilities at Paranal Observatory under program IDs 091.D-0020 and 093.D-0316.}\fnmsep\thanks{Based on observations collected at ESO La Silla Observatory using the Coralie spectrograph mounted to the Swiss 1.2 m Euler telescope, under program CNTAC2014A-5.}}

   \author{J. Breitfelder\inst{1, 2}
          \and
          P. Kervella\inst{2, 3}
          \and
          A. M\'{e}rand\inst{1}
          \and
          A. Gallenne\inst{4}
	\and
	L. Szabados\inst{5}
	\and
	R. I. Anderson\inst{6}
	\and
	M. Willson\inst{7}
	\and
	J.-B. Le Bouquin\inst{8}
         }

   \institute{European Southern Observatory, Alonso de C\'{o}rdova 3107, Casilla 19001, Santiago 19, Chile\\
         \email{joanne.breitfelder@obspm.fr}
         \and
         LESIA, Observatoire de Paris, CNRS UMR 8109, UPMC, Universit\'{e} Paris Diderot, 5 place Jules Janssen, 92195 Meudon, France
         \and
	Unidad Mixta Internacional Franco-Chilena de Astronom\'{i}a, CNRS/INSU, France (UMI 3386) and
	Departamento de Astronom\'{i}a, Universidad de Chile, Camino El Observatorio 1515, Las Condes, Santiago, Chile	
	\and
         	Universidad de Concepci\'{o}n, Departamento de Astronom\'{i}a, Casilla 160-C, Concepci\'{o}n, Chile
	\and 
	Konkoly Observatory of the Hungarian Academy of Sciences, H-1121 Budapest, Konkoly Thege Str. 15-17, Hungary
	\and 
	Observatoire de Gen\`eve, Universit\'e de Gen\`eve, 51 Ch. des Maillettes, 1290 Sauverny, Switzerland
	\and
	School of Physics and Astronomy, University of Exeter, Stocker Road, Exeter EX4 4QL, UK
	\and
	UJF-Grenoble 1 / CNRS-INSU, Institut de Plan\'{e}tologie et d'Astrophysique de Grenoble (IPAG) UMR 5274, Grenoble, France
          }

   \date{Received date / Accepted date}

  \abstract
   {The distances of pulsating stars, in particular Cepheids, are commonly measured using the parallax of pulsation technique. The different versions of this technique combine measurements of the linear diameter variation (from spectroscopy) and the angular diameter variation (from photometry or interferometry) amplitudes, to retrieve the distance in a quasi-geometrical way.
   However, the linear diameter amplitude is directly proportional to the projection factor (hereafter \emph{p}-factor), which is used to convert spectroscopic radial velocities (i.e., disk integrated) into pulsating (i.e., photospheric) velocities. The value of the $p$-factor and its possible dependence on the pulsation period are still widely debated.}
   {Our goal is to measure an observational value of the $p$-factor of the type-II Cepheid $\kappa$~Pavonis.}
   {The parallax of the type-II Cepheid $\kappa$~Pav was measured with an accuracy of 5\% using HST/FGS. We used this parallax as a starting point to derive the $p$-factor of $\kappa$~Pav, using the {\tt SPIPS} technique (Spectro-Photo-Interferometry of Pulsating Stars), which is a robust version of the parallax-of-pulsation method that employs radial velocity, interferometric and photometric data. We applied this technique to a combination of new VLTI/PIONIER optical interferometric angular diameters, new CORALIE and HARPS radial velocities, as well as multi-colour photometry and radial velocities from the literature.}
   {We obtain a value of $p = \pfactor \pm \pfactorerrtot$ for the \emph{p}-factor of $\kappa$~Pav. This result agrees with several of the recently derived Period-$p$-factor relationships from the literature, as well as previous observational determinations for Cepheids.}
   {Individual estimates of the \emph{p}-factor are fundamental to calibrating the parallax of pulsation distances of Cepheids. Together with previous observational estimates, the projection factor we obtain points to a weak dependence of the $p$-factor on period.}


\maketitle
%

\section{Introduction}
\label{kappapav}


Cepheids have been used for more than a century as standard candles for estimating extragalactic distances, owing to the linear relationship between the logarithm of their pulsation period and their intrinsic magnitude: $M = a\, \left(\log{P} - 1\right) + b$, where $P$ is the period of pulsation and $M$ the mean absolute magnitude (hereafter the P-L relation).
This remarkable relation was discovered empirically in 1908 by Henrietta Leavitt \citep{1908AnHar..60...87L, 1912HarCi.173....1L} and is now named after her : \emph{Leavitt's Law}. While the slope $a$ of the P-L relation can be estimated by observing a number of Cepheids located, for example, in the Magellanic Clouds, the zero point $b$ is more difficult to estimate. It was first settled by \citet{1913AN....196..201H}, when estimating the LMC distance. Thanks to the small dispersion of Leavitt's Law, Cepheids have, for more than a century, been considered as one of the most accurate standard candles for estimating extragalactic distances. As a recent example, Cepheids were involved in the $3\%$ measurement of the Hubble constant $H_0$ presented by \citet{2011ApJ...730..119R}, as calibrators of the magnitude-redshift relation of type Ia supernovae. It should be noted that half of the $3\%$ error budget is attributed to the uncertainty on the Cepheid distance scale. The uncertain calibration of the P-L relation, as well as photometric biases (e.g., from reddening) can introduce important systematic uncertainties. Leavitt's Law is also affected by physical effects like metallicity. A better measurement of $H_{0}$ is fundamental to constraining the $\Lambda CDM$ model parameters, such as the number of neutrino species and the equation of state of the dark energy $\omega_c$. A detailed review of the methods used to determine the history of the expansion and their current limitations is given in \citet{2013PhR...530...87W}. Better precision would also help for addressing the 2$\sigma$ tension between the value of \citet{2011ApJ...730..119R} and the value derived from \emph{Planck}'s CMB modelling \citep{2014A&A...566A..54P}, and revealing potential biases in one of these methods. 

A better estimate of $H_{0}$ requires a more accurate calibration of the P-L relationship \citep{2012arXiv1202.4459S}. Parallax measurements can be used to determine the zero point of the P-L relationship, but they are reasonably accurate only for a few nearby Cepheids. Other distance estimates, such as the Baade-Wesselink (hereafter BW) technique \citep{2013IAUS..289..138G} and its infrared surface-brigthness version \citep{2011A&A...534A..94S}, the light echoes of RS Pup \citep{2014arXiv1408.1697K}, or binary Cepheid orbital parallax \citep{2013MNRAS.436..953P, 2013A&A...552A..21G, 2014A&A...561L...3G}, should be obtained as independent cross-checks to ensure the accuracy of the calibration.
We are currently carrying out a long-term programme of interferometric observations of Cepheids in both hemispheres, using the CHARA array installed at the Mount Wilson Observatory in California \citep{2005ApJ...628..453T} and the Very Large Telescope Interferometer (hereafter VLTI) installed at the Cerro Paranal in Chile \citep{2014arXiv1407.2785M}. This programme has led to the discovery of circumstellar envelopes around several Cepheids \citep[e.g.][]{2012A&A...538A..24G, 2009A&A...498..425K} and companions \citep[e.g.][]{2013A&A...552A..21G}.

The BW technique \citep{1926AN....228..359B} combines measurements of the linear diameter variation (from spectroscopy) and the angular diameter variation (from either photometry or interferometry) to retrieve the distance in a quasi-geometrical way. The accuracy of this elegant method is although limited by the projection factor (hereafter $p$-factor) used to convert spectroscopic radial velocities into pulsating velocities representing the actual displacement of the photosphere. The value of the $p$-factor, and its possible dependence on the pulsation period, are still debated. 
We developed a dedicated software tool to estimate the distance of pulsating Cepheids, the \texttt{SPIPS} code \citep{2013IAUS..289..183M} that we briefly present in Sect.~\ref{algo}. Taking advantage of the distance to $\kappa$~Pavonis obtained by \citet{2011AJ....142..187B} using an independent method, we track down in the present work the $p$-factor of this star through an inverse application of this algorithm (Sect.~\ref{pfactor}). 


\section{Observations and data processing \label{dataprocessing}}

\subsection{The peculiar Cepheid $\kappa$~Pavonis}
\label{kappapav}

As one of the closest type-II Cepheids \citep{2002PASP..114..689W}, $\kappa$~Pav is classified as a member of the W Vir class \citep{1963MNRAS.125..487R}, which groups pulsators with periods between 10 and 20 days. Since it is slightly brighter and bluer than the normal population of this class, \citet{2009MNRAS.397..933M} propose to classify it as a pW star, which is a new category introduced by \citet{2008AcA....58..293S} to describe those peculiar W Vir Cepheids. Type-II Cepheids are luminous stars, with similar behaviour to classical Cepheids and variation periods in the same range. But their lower mass makes them representatives of the population II stars present in the galactic halo and the old galactic disk. $\kappa$~Pav also shows similarities with RR Lyrae, placing it at an interesting interface between classical Cepheids, type-II Cepheids and RR Lyrae pulsators.

\subsection{VLTI/PIONIER long-baseline interferometry}

Interferometric angular diameter measurements of $\kappa$~Pavonis were carried out in August and September 2013 and from June to August 2014 using the four-telescope beam combiner PIONIER, installed at the VLT Interferometer in northern Chile \citep{2010SPIE.7734E..99B, 2011A&A...535A..67L}. The observations were undertaken in three spectral channels of the $H$ band (1.59 $\mu$m, 1.67 $\mu$m, and 1.76 $\mu$m), corresponding to a low spectral resolution of $R = 40$. We used the four relocatable 1.8-metre Auxiliary Telescopes (ATs), installed at stations A1-G1-J3-K0\footnote{https://www.eso.org/paranal/telescopes/vlti/configuration/}. This configuration offers the longest available baselines (from 57 meters between the stations K0 and J3, up to 140 metres between A1 and J3) that are required to get the necessary angular resolution to resolve the apparent disk of $\kappa$~Pav ($\theta \sim 1.2$ mas).

\begin{figure}
	\resizebox{\hsize}{!}{\includegraphics{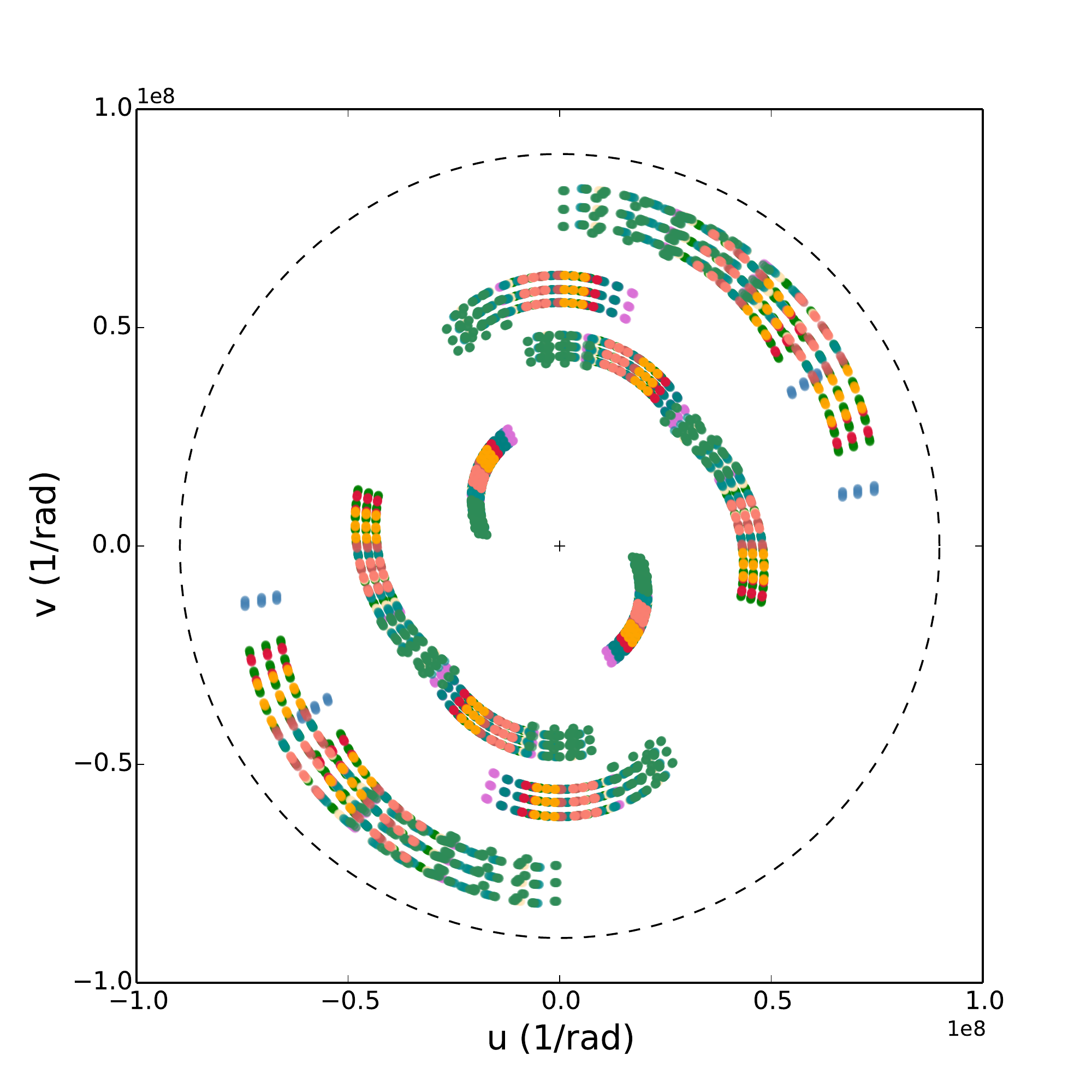}}
	\caption{(u,v) coverage of our PIONIER observations (each colour corresponds to one night of observations). The black dashed-line circle traces the spatial frequency where the squared visibility is equal to 0.5, assuming for $\kappa$~Pav an average diameter of $\theta_{Avg.} = 1.182$ mas (resulting from the present study).}
	\label{<uvmap>}
\end{figure}

Figure~\ref{<uvmap>} shows a map of the $(u,v)$ plane coverage of our observations. We obtained a very good sampling of the pulsation curve, including data points close to the maximum and the minimum diameters. That is particularly interesting for the present work, because the most precise diameter variation is needed to derive an accurate value of the \emph{p}-factor. The most suitable calibrators have been chosen from the catalogue of \citet{2005A&A...433.1155M}. Their main characteristics are given in Table~\ref{table:1}.

\begin{figure}
	\resizebox{\hsize}{!}{\includegraphics{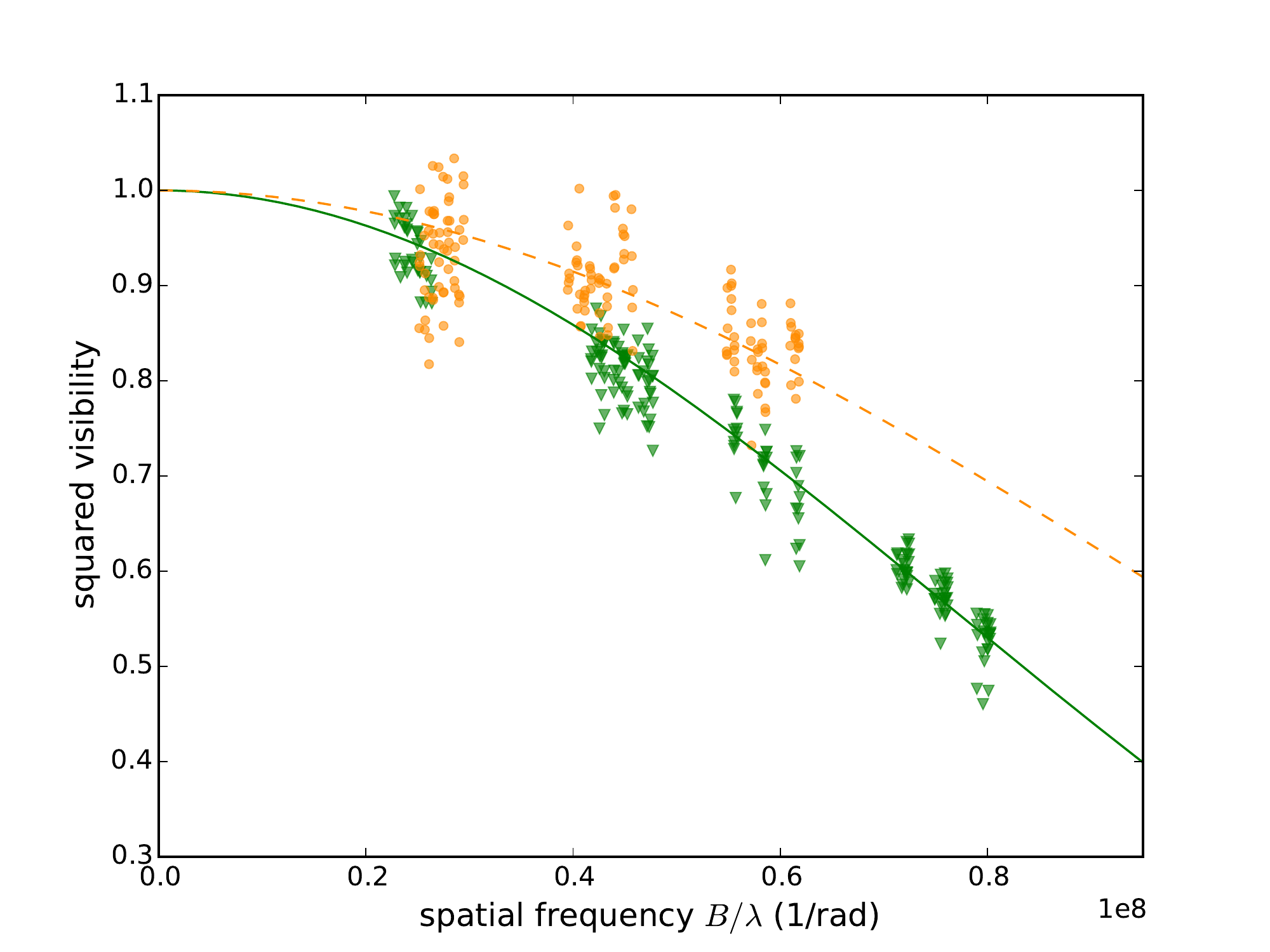}}
	\caption{Squared visibilities measured with PIONIER at the minimum and the maximum diameters. The data are fitted with a uniform disk model leading to the diameters of $\theta_{min}=0.976$ mas (MJD=56 539, in yellow) and $\theta_{max}=1.273$ mas (MJD=56 871, in green). Errorbars have been removed for clarity.}
	\label{<visibilities>}
\end{figure}

\begin{table}
	\caption{Main caracteristics of the calibrators used during our PIONIER observations.}
	\label{table:1}
	\centering
	\renewcommand{\arraystretch}{1.2}
	\begin{tabular}{*{5}{c}}
		\hline
		Star 			& $m_{v}$ & $m_{H}$ & $\theta_{UD} \pm \sigma$ (mas) \\ 
		\hline
		\hline 
		HD 181019 	& 6.35 	& 3.31 	& 1.030 $\pm$ 0.014 \\
		HD 171042 	& 7.51 	& 3.63 	& 1.030 $\pm$ 0.014 \\
		HD 175782 	& 6.80 	& 3.05 	& 1.337 $\pm$ 0.019 \\
		HD 186530 	& 7.41 	& 4.01 	& 0.896 $\pm$ 0.012 \\
		\hline
	\end{tabular}
	\tablefoot{Calibrators were selected from the catalogue of \citet{2005A&A...433.1155M}. We indicate the uniform disk diameter in H band and the corresponding uncertainty.}
\end{table}

The raw data were reduced using the {\tt pndrs} data reduction software of PIONIER \citep{2011A&A...535A..67L}, which produces calibrated squared visibilities and phase closures. The resulting {\tt OIFITS} format of the data allowed us to compute a uniform disk (UD) value for each observing epoch with the model fitting software \texttt{LITPro}\footnote{The LITpro software is available at http://www.jmmc.fr/litpro} \citep{2008SPIE.7013E..44T}. The results are listed in Table~\ref{table:2}. For each night of observations we give the mean Julian date, the phase, the UD diameter, and its uncertainty. The error given for each diameter includes a statistical error given by the model fitting software and a systematic one (which actually dominates the error bugdet), because of the uncertainty on the calibrator diameters. We therefore define this systematic error as being the mean of the errors of all the calibrators. The phases given in Table~\ref{table:2} were computed by using the ephemeris derived from the O-C diagram shown in Fig.~\ref{<OCdiag>} (see details in Sect.~\ref{phasing}). The variations in the diameter of the Cepheid can be appreciated in Fig.~\ref{<visibilities>}, which represents the PIONIER squared visibilities and the corresponding best UD model for the maximum and minimum diameters.

\begin{table}
	\caption{PIONIER observations of $\kappa$~Pav. We give here the mean MJD (defined by JD$-$2\,400\,000.5) of each observing night, the corresponding phase (taking $\phi_{0}$ at the maximum of luminosity), the best uniform disk diameter adjusted on the squared visibility measurements and its uncertainty. We also give the $\chi^2$ resulting from the fit of the squared visibilities with a uniform disk model.}
	\label{table:2}
	\centering
	\renewcommand{\arraystretch}{1.2}
	\begin{tabular}{*{4}{c}}
		\hline
		MJD 		& Phase  & $\theta_{UD} \pm \sigma_{\rm{stat.}} \pm \sigma_{\rm{syst.}}$ (mas) & $\chi^2$ \\
		\hline
		\hline
		56508.024 	& 0.375	& 1.256 $\pm$ 0.004 $\pm$ 0.014 & 0.96 \\
		56509.009 	& 0.483 	& 1.243 $\pm$ 0.002 $\pm$ 0.015 & 3.30 \\
		56538.065 	& 0.682 	& 1.101 $\pm$ 0.012 $\pm$ 0.016 & 1.30 \\   
		56539.046 	& 0.790	& 0.976 $\pm$ 0.016 $\pm$ 0.016 & 1.35 \\
		56540.020 	& 0.898 	& 1.159 $\pm$ 0.012 $\pm$ 0.015 & 0.59 \\
		56831.126 	& 0.948 	& 1.101 $\pm$ 0.004 $\pm$ 0.014 & 2.21 \\
		56833.166 	& 0.173 	& 1.262 $\pm$ 0.002 $\pm$ 0.014 & 1.73 \\
		56866.992	& 0.897	& 1.057 $\pm$ 0.002 $\pm$ 0.014 & 1.01 \\
		56869.122	& 0.132	& 1.222 $\pm$ 0.007 $\pm$ 0.014 & 1.48 \\
		56871.055	& 0.344	& 1.273 $\pm$ 0.004 $\pm$ 0.014 & 0.94 \\
		56874.090	& 0.679	& 1.111 $\pm$ 0.004 $\pm$ 0.014 & 1.10 \\
		56894.017	& 0.872	& 1.026 $\pm$ 0.001 $\pm$ 0.014 & 0.69 \\
		\hline
	\end{tabular}
\end{table}

\subsection{Period changes, overall phasing, and photometry} \label{phasing}

\begin{figure}
	\resizebox{\hsize}{!}{\includegraphics{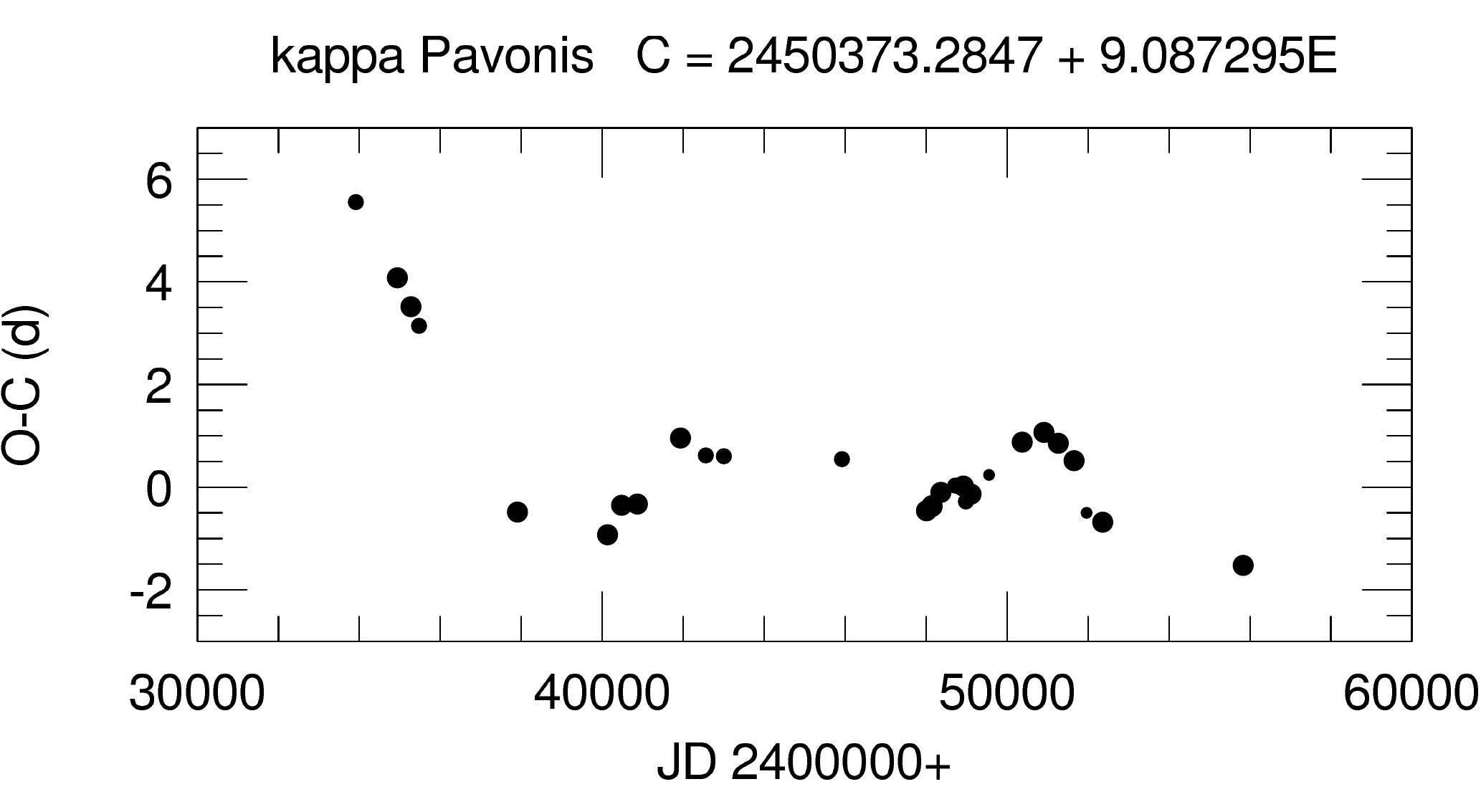}}
	\caption{O-C diagram of $\kappa$~Pav. A weighted least squares fit to the residuals leads to $T_{0}=2450373.2847$ and $P=9.0873$. Considering only the data with JD>2450000, we get $T_{0}=2450374.2938$ and $P=9.0827$.}
	\label{<OCdiag>}
\end{figure}

The pulsation period of $\kappa$~Pav shows large and fast variations that complicate the phasing of datasets from different epochs. By applying the statistical Eddington-Plakidis method \citep{1929MNRAS..90...65E} on a large photometric dataset, \citet{2009SPIE.7651E..31B} showed that the period variations of $\kappa$~Pav are erratic, which has already been suggested in previous studies \citep[e.g.][]{2008MNRAS.386.2115F, 1992MNRAS.259..474W}, with a relatively high degree of randomness. To help the phasing of the data used in the present study, we computed ephemerides from the O-C diagram shown in Fig.~\ref{<OCdiag>}.  When constructing the O-C diagram, only photoelectric and CCD photometric data have been taken into account. Depending on the number of observations and phase coverage of the individual datasets, a weight of 1, 2, or 3 has been assigned to the derived moment of the normal maximum brightness. In Fig.~\ref{<OCdiag>}, the size of the filled circles refers to the weight assigned to the given residual (O-C difference). A weighted least squares fit of the data for JD>2440000 leads to $T_{0}=2450373.2847$ and $P=9.0873$ days. Considering only the data for JD>2450000, we get $T_{0}=2450374.2938$ and $P=9.0827$ days. These values have been used to phase our most recent data.

We selected the following photometric data from the literature: \emph{Hipparcos} and \emph{Tycho} photometry from the \citet{1997ESASP1200.....E} catalogue (see also \cite{1997A&A...323L..61V}), $JHK$ photometry from \citet{2008MNRAS.386.2115F} (hereafter F08), $VBLUW$ photometry from \citet{1964BAN....17..520W}, and $UBV R_c I_c$ photometry from \citet{2008yCat.2285....0B}. We divided this last dataset into three groups covering different epochs of around 700 days and separated by around 350 days (Group 1: MJD from 50347 to 50917, Group 2: 51248 to 51972 and Group 3: 52323 to 53118). We then phased the resulting photometric sequences separately.
Considering the erratic changes in the period of $\kappa$~Pav, we computed specific $(P,T_0)$ ephemeris for each dataset. We used the \texttt{SPIPS} code to derive a reference epoch and the corresponding pulsation period. In the case of the \emph{Hipparcos} and \emph{Tycho} data, we introduced a linear variation in the period, which allowed us to reach a more satisfying phasing. For each photometric dataset, the reference epoch was set at the maximum luminosity of the star, and taken as close as possible to the centre of the epoch range. The resulting periods and reference epochs $\textrm{MJD}_{0}$ are given in Table~\ref{table:3} and were used to compute the phases of the observations. 

We note that the \emph{Hipparcos} and \emph{Tycho} data seem to be relatively dispersed around the phase 0.6, given their low errorbars. This could be due to possible amplitude variations, which have already been observed in the photometry of type II Cepheids and overtone pulsators \citep{2014arXiv1411.1730E}. However, this does not affect the results of the present study.

Table~\ref{table:3} also sums up the method used to phase the other observables : for the radial velocities from \citet{1992MNRAS.259..474W} (see details in Sect. \ref{RVs}), we kept the phases provided by the author and shifted the whole curve to make it match the photometry. Our CORALIE and HARPS data were phased by using the ephemeris derived from the O-C diagram (see Fig.~\ref{<OCdiag>}), and then shifted into phase agreement with the rest of the data. We used the same ephemeris for our PIONIER diameters, but we did not need to add any phase shift.

\begin{table*}
	\caption{Ephemerides used to phase the interferometric, spectroscopic, and photometric data.}
	\label{table:3}
	\centering
	\renewcommand{\arraystretch}{1.2}
	\begin{tabular}{*{5}{c}}
		\hline
		Data reference 					& Phase shift 		& $\textrm{MJD}_{0}$ 	& Period (days) & Period drift (s/year) \\
		\hline
		\hline
		\multicolumn{5}{c}{\emph{Radial velocities}}\\
		\cite{1992MNRAS.259..474W}		& $\phi + 0.171$	&	-				&	-		&	-	\\
		Present work						& $\phi + 0.034$	& 50373.793			& 9.0827		&	-	\\
		\hline
		\multicolumn{5}{c}{\emph{Interferometry}}\\
		Present work						&	-			& 50373.793			& 9.0827		&	-	\\
		\hline	
		\multicolumn{5}{c}{\emph{Photometry}}\\
		\cite{2008yCat.2285....0B}	(1)		&	-			& 50646.541			& 9.0897		&	-	\\
		\cite{2008yCat.2285....0B}	(2)		&	-			& 51609.425			& 9.0804		&	-	\\
		\cite{2008yCat.2285....0B}	(3)		&	-			& 52353.321			& 9.0669		&	-	\\
		\cite{2008MNRAS.386.2115F}		&	-			& 46800.743			& 9.0789		&	-	\\
		\cite{1964BAN....17..520W}			&	-			& 37913.807			& 9.0833		&	-	\\
		\cite{1997ESASP1200.....E}		&	-			& 47482.679			& 9.0888		& 65.895 \\
		\hline
	\end{tabular}
\end{table*}

\subsection{Radial velocities} \label{RVs}

\begin{table}
	\caption{Radial velocity measurements deduced from the spectra obtained with CORALIE and HARPS, both installed in La Silla observatory in Chile. We adopted a total error budget of 500 m/s for each observation.}
	\label{table:5}
	\centering
	\renewcommand{\arraystretch}{1.2}
	\begin{tabular}{*{5}{c}}
		\hline
		MJD 		& Phase 	& $V_{\rm{rad}} \pm \sigma_{\rm{tot.}} $ (km/s) 	& Instrument\\
		\hline
		\hline
		56750.416 	& 0.096	& 28.74 $\pm$ 0.50   & CORALIE \\
		56751.412	& 0.206	& 33.32 $\pm$ 0.50   & CORALIE \\
		56826.428	& 0.465	& 44.54 $\pm$ 0.50   & CORALIE \\
		56827.425	& 0.575	& 49.31 $\pm$ 0.50   & CORALIE \\
		
		56876.174	& 0.942	& 26.09 $\pm$ 0.50   & HARPS \\
		56876.976	& 0.031	& 27.55 $\pm$ 0.50   & HARPS \\
		56877.974	& 0.140	& 31.77 $\pm$ 0.50   & HARPS \\
		56878.975	& 0.251	& 36.26 $\pm$ 0.50   & HARPS \\
		56907.984	& 0.444	& 44.13 $\pm$ 0.50   & HARPS \\
		56908.983	& 0.554	& 48.92 $\pm$ 0.50   & HARPS \\
		56909.994	& 0.666	& 51.72 $\pm$ 0.50   & HARPS \\
		\hline
	\end{tabular}
\end{table}

We retrieved three sets of radial velocity measurements from the literature, resulting from observations carried out between 1904 and 1918 \citep{1929LicOB..14...60J}, in 1961 \citep{1963MNRAS.125..487R} and in 1988 \citep{1992MNRAS.259..474W} (hereafter W92). As the data from \citet{1929LicOB..14...60J} and \citet{1963MNRAS.125..487R} show a relatively high dispersion, we only used the metallic-line radial velocities from W92. The convention used to derive the phases given in W92 is uncertain (Wallerstein, private communication) and by computing our own ephemerides, we did not succeed in obtaining a radial velocity curve that was as well-phased as the one given in W92. We therefore decided to keep the phases given in the paper. Because the zero-phase definition in W92 does not correspond to the maximum flux, we shifted the original phase values to match the zero-phase convention of the photometry.

We also obtained contemporaneous and very precise radial velocities (RVs) between November 2013 and June 2014 using the high-resolution Echelle spectrographs HARPS mounted on the ESO 3.6m telescope and CORALIE mounted on the Swiss 1.2m Euler telescope, both of which are located at ESO's La Silla observatory in Chile. The RVs were derived using the cross-correlation technique \citep{1996A&AS..119..373B, 2002A&A...388..632P} and a standard Gaussian fit.

Unfortunately, the new measurements are not sufficient to determine a very precise pulsation period. Therefore, we computed the phases of
the RV measurements using the ephemeris used for the PIONIER data. We then shifted the whole curve by adding a constant in phase, until obtaining the lowest reduced $\chi^2$ at the end of the fitting process. The phasing process is summarized in Table~\ref{table:3}. While we took great care to obtain the best possible phasing for the new RV data, there is a remaining uncertainty regarding the phases of the new measurements. This problem is exacerbated by known random period fluctuations (see details given in section~\ref{phasing}). To reduce the sensitivity of our method to phase errors of the new RV data (this easily incurs errors of several 100 m/s and thus dominates the uncertainty of the RV curve), we attribute a reduced weight to the new RV data by adopting a fixed error budget of 500 m/s in the fit. This does not constitute a limiting factor for our results, since the accuracy of our $p$-factor determination is limited by the parallax accuracy of 5\% (see details given in section~\ref{pfactor}).

\section{The \texttt{SPIPS} algorithm \label{spips}}

\subsection{The parallax of pulsation method and the $p$-factor limitation}
\label{parallaxofpulsation}

The fundamental equation of the parallax of pulsation method can be written as follows, where $\theta_{t=0}$ is the angular diameter at the maximum of luminosity, $d$ the distance and $v_{\textrm{puls}}$ the pulsation velocity of the atmosphere:
\begin{equation}
\theta(t) = \theta_{t=0} + \frac{2}{d} \int_{0}^{t} v_{\textrm{puls}}(\tau)d\tau 
\end{equation}

The main limitation of the parallax of pulsation technique comes from the projection factor \emph{p} used to convert the radial velocities from the Doppler shift of spectral lines (a disk-integrated quantity) into pulsation velocities (the displacement velocity of the photosphere over the pulsation cycle):
\begin{equation}
v_{\textrm{puls}} = p\,v_{\textrm{rad}} 
\end{equation}

We do not consider here the effects of amplitude modulations as recently reported for Cepheids by \citet{2014A&A...566L..10A}. We can therefore rewrite the main equation of the parallax of pulsation as follows:

\begin{equation} \label{pfactoreq}
\theta(t) = \frac{2}{d} \left( R_{0} + \int_{0}^{t} p\,v_{\textrm{rad}}(\tau)d\tau \right)
\end{equation}

Only the projected component of the velocity on the line of sight contributes to the measured Doppler shift of the spectral lines. The spherical geometry of the star results in a projection effect corresponding to a value of $p=1.5$ for a uniform brightness sphere. The limb darkening and the dynamical behaviour of the line-forming regions in the stellar atmosphere are expected to reduce the value of $p$ below this value \citep{2007A&A...471..661N}. Unfortunately, these effects can hardly be quantified separately. 

The combination of these different effects results in a relatively large dispersion of the \emph{p}-factor estimates found in the literature. Some recent studies propose either a constant value \citep{2008MNRAS.386.2115F, 2007A&A...474..975G, 2005A&A...438L...9M} or Period-$p$-factor relationships \citep{2013A&A...550A..70G, 2012Ap&SS.341..115S, 2012A&A...543A..55N, 2009A&A...502..951N}. As a consequence of this uncertainty, the truly unbiased quantity that can be derived using the parallax of pulsation method is the ratio $d/p$, where $d$ is the distance (Eq.~\ref{pfactoreq}). Observational determinations of $p$ are therefore critical for better constraining the \emph{p}-factor models and eventually establishing the parallax of pulsation method on solid foundations. A review of the role and importance of the $p$-factor can be found in \citet{2014IAUS..301..145N}.

The observational calibration of the $p$-factor requires independent measurements of the distance $d$ for a sample of Cepheids, to waive the degeneracy of the $d/p$ ratio produced by the parallax of pulsation method. In this case, when the distance is known (e.g. from trigonometric parallax, light echoes, or binary Cepheids), the inversion of the method gives access to the \emph{p}-factor. This has already been done on the classical Cepheid $\delta$ Cephei by \citet{2005A&A...438L...9M}, yielding $p = 1.27 \pm 0.06$. We apply here this method to $\kappa$~Pav using the enhanced \texttt{SPIPS} algorithm (Sect.~\ref{spips}) and considering the accurate parallax recently measured by \citet{2011AJ....142..187B} using the HST/FGS interferometer ($\pi = 5.57 \pm 0.28$\,mas).

\subsection{Overview of the algorithm}
\label{algo}

Classical BW methods are limited by various systematic errors (e.g. photometry biases or reddening) that affect either the photometry or the spectroscopic observables. To overcome this, we developed a dedicated tool (\texttt{SPIPS} for Spectro-Photo-Interferometry of Pulsating Stars, \cite{2013IAUS..289..183M, 2014IAUS..301..389B}; M\'{e}rand et al., \emph{in prep.}) that computes a global fit of all the available data (i.e., radial velocities, interferometric squared visibilities, spectroscopically determined effective temperatures ($T_{\rm{eff}}$), colours and magnitudes in various bands and filters).
This combination of different observables allows us to reach a better accuracy on our measurements (2 to $5\,\%$ uncertainty on the distance for an individual Cepheid). The partial redundancy of the data (e.g., interferometry and atmospheric models applied to the photometry to estimate the angular diameter) results in much improved robustness of the fitting process. Besides that, the integration of physical models in our code (e.g. \texttt{ATLAS9} atmospherical models from \cite{1979ApJS...40....1K}) reduces the statistical biases (owing to the calibration of the zero point of the photometric filters, for instance). The larger overall amount of observational data also reduces the statistical errors on the resulting parameters (e.g., the distance, the colour excess $E(B-V)$, and the mean angular diameter and effective temperature). The $\chi^2$ minimization process is optimized by defining the global $\chi^2$ as the average of the specific reduced $\chi^2$ values for each data set. This allows us to adjust the relative weight of each type of data in the global fit. Otherwise, in the present case, the diameter and the radial velocity would not contribute as much as the photometry in the minimization process.

\subsection{Projection factor}
\label{pfactor}

\begin{figure} 
	\includegraphics[width=\hsize]{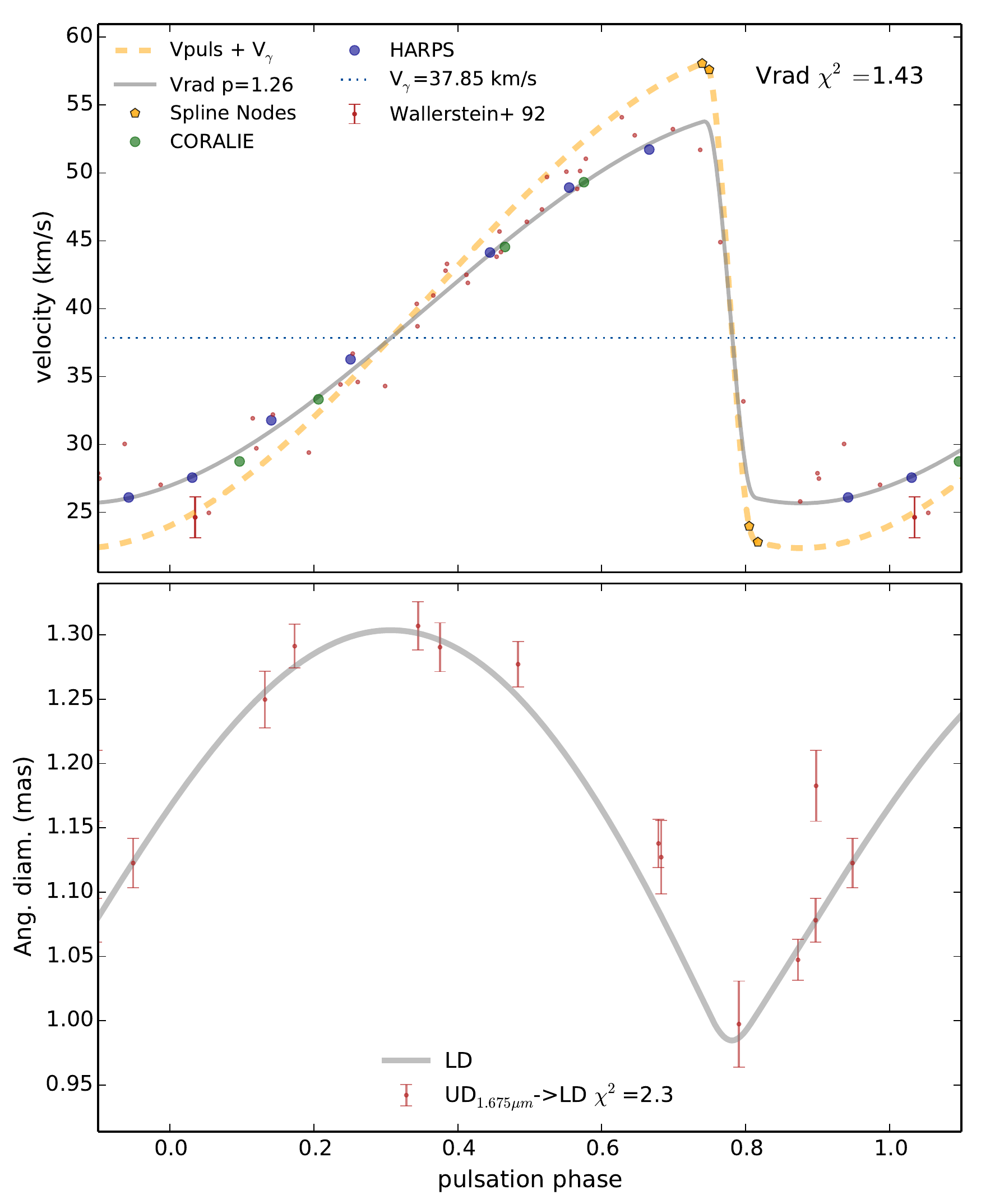}
	\caption{\texttt{SPIPS} code applied on the type-II Cepheid $\kappa$~Pav. \emph{Above:} Radial velocities from \citet{1992MNRAS.259..474W} (the typical error bar is shown in the bottom left corner) and new measurement from CORALIE and HARPS (the size of the points corresponds to the errorbar), fitted using spline functions. \emph{Below:} Uniform disk angular diameters deduced from our PIONIER interferometric observations. For each panel we indicate the reference and the individual reduced $\chi^2$.}
	\label{<diam_vrad>}
\end{figure}

\begin{figure} 
	\includegraphics[width=\hsize]{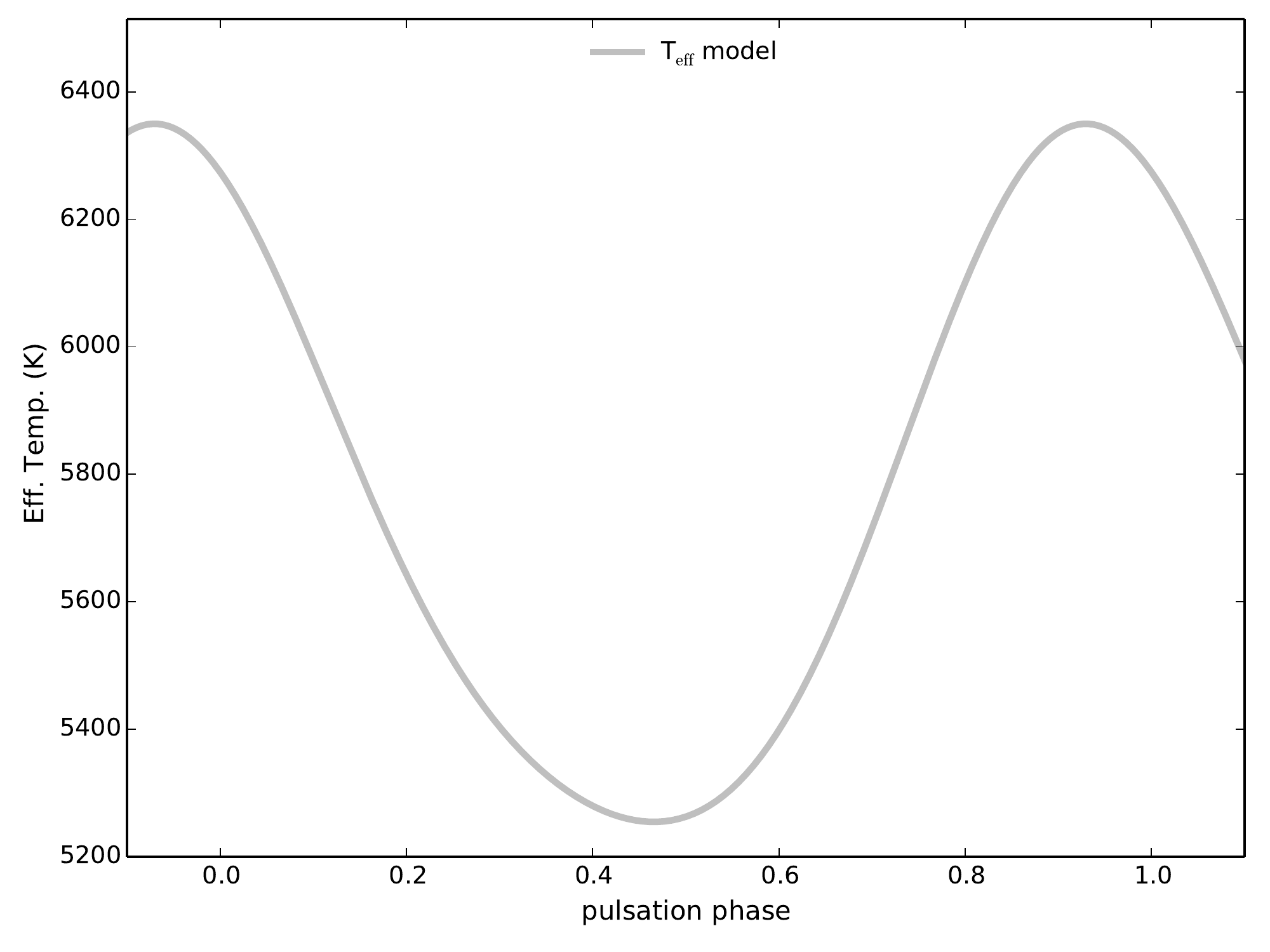}
	\caption{\texttt{SPIPS} code applied on the type-II Cepheid $\kappa$~Pav. Effective temperature curve deduced from \texttt{ATLAS9} atmospheric model grids \citep{2004astro.ph..5087C, 2005MSAIS...8...14K}}
	\label{<Teff>}
\end{figure}

\begin{figure*} 
	\includegraphics[width=\hsize]{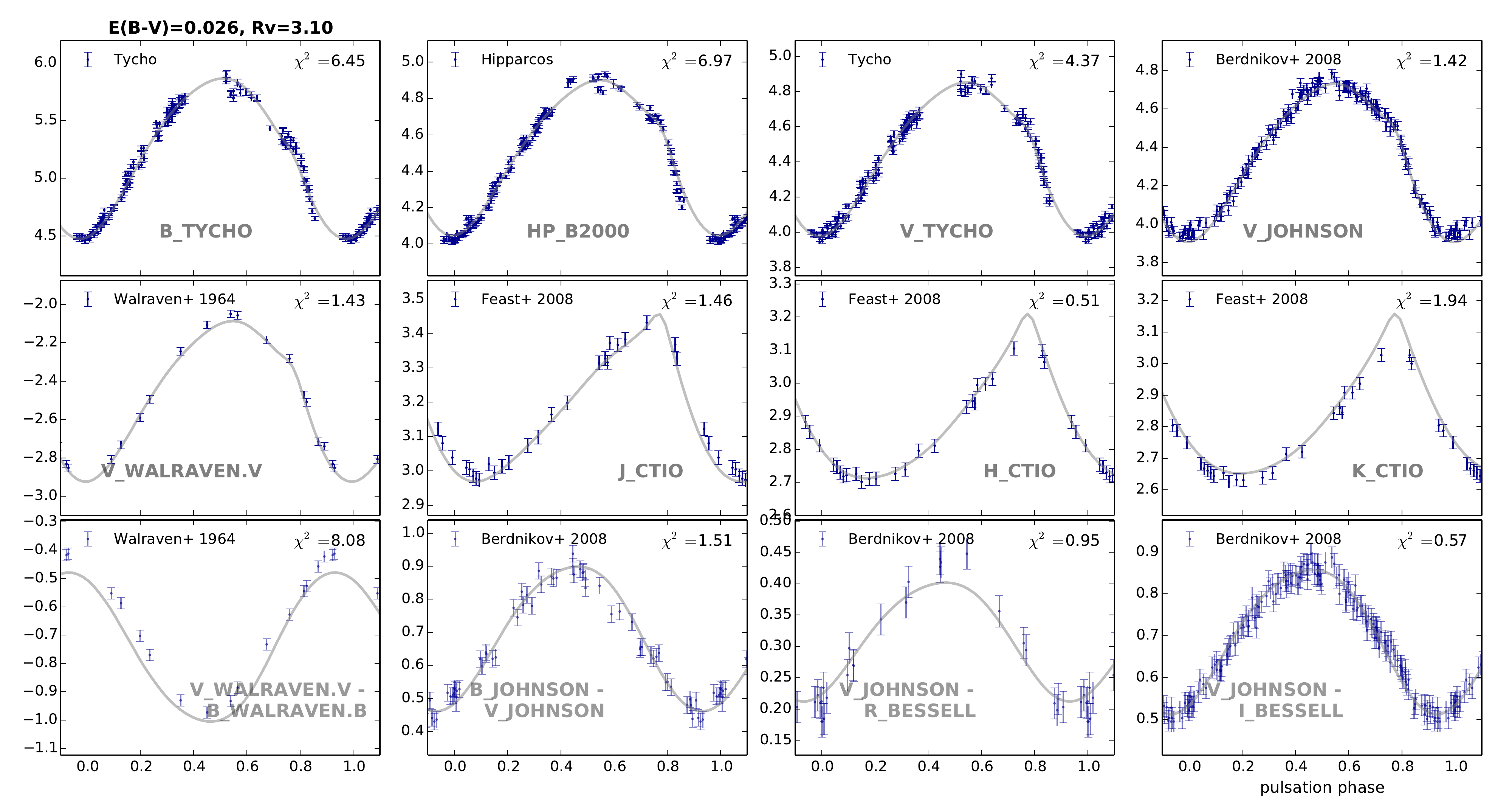}
	\caption{\texttt{SPIPS} code applied on the type-II Cepheid $\kappa$~Pav : magnitudes and colours fitted with Fourier series. For each dataset we precise the filter, the reference, and the individual reduced $\chi^2$.}
	\label{<photom>}
\end{figure*}

We fitted the radial velocity with spline functions defined by four adjustable floating nodes. Although it is numerically less stable than Fourier series, it is necessary to reproduce the strong slope of the radial velocity curve that occurs between the phases 0.8 and 1.0 without introducing high frequency variations in the rest of the curve. 
The photometry curve is adjusted to the data using a second-order Fourier series. 
We shifted the $(B-R)$ and $(B-I)$ colours from \citet{2008yCat.2285....0B} vertically, by subtracting 0.008 magnitudes to $(B-I)$ and 0.038 magnitudes to $(B-R)$. Otherwise, an offset always remained between the data and the model, probably because of a bias introduced by the calibration of the zero point of the filter used for the observations in $I$ and $R$. It is then important to note that only the shape of these two particular curves continues to be constraining.
To give an equivalent weight to the different observables in the fitting process, we multiplied the errors by the normalization factors (NF) given in Table~\ref{table:6}. These coefficients depend on the number of points ($N_{points}$) in each dataset. Table~\ref{table:6} also gives the individual reduced $\chi^2$ of the different datasets. 

We fixed the metallicity at $[\rm{Fe}/\rm{H}]=0.0$ \citep{1989ApJ...342..476L} and the distance at $d=179$\,pc \citep{2011AJ....142..187B}. The best-fit parameters are given in Table~\ref{table:4}. The values and statistical uncertainties have been determined through a Monte-Carlo procedure. We obtain for $\kappa$~Pav a \emph{p}-factor $p = \pfactor \pm \pfactorerrstat_{\rm{stat.}} \pm \pfactorerrsyst_{\rm{syst.}}$, with a systematic error given by the limited 5\% precision on the distance. The final value is therefore $p = \pfactor \pm \pfactorerrtot$. The systematic error on the average $T_{\rm{eff}}$ has been deduced from the Stefan-Boltzmann law, after considering a systematic error of 5\% on the photometry and of 1.4\% on the angular diameter (given by the limited precision of the calibrators diameters). The systematic uncertainty on the reddening has been estimated by computing the maximum and minimum values of the colour index $(B-V)_{0}$ obtained in the uncertainty range of the effective temperature \citep{1996ApJ...469..355F}. 

\begin{table}
	\caption{Best-fit output parameters given by the \texttt{SPIPS} code.}
	\label{table:4}
	\centering
	\renewcommand{\arraystretch}{1.2}
	\begin{tabular}{llll}
		\hline
		Parameter					& Value 		& $\sigma_{\rm{stat.}}$	& $\sigma_{\rm{sys.}}$	\\
		\hline
		\hline
		\emph{p}-factor 				& \pfactor  	& \pfactorerrstat		&	\pfactorerrsyst		\\
		$\theta$ (UD) at $\phi = 0$ (mas) 	& 1.1654		& 0.0025				&	0.014			\\
		Avg. $\theta$ (UD) (mas)			& 1.1823		& 0.0021				&	0.014			\\
		$\rm{v}_{\gamma}$ (km/s)		& 37.87		& 0.18				&	0.50				\\
		$E(B-V)$ (mag)				& 0.02		& 0.01				&	0.04				\\
		Avg. $T_{\rm{eff}}$ (K)			& 5739		& 9					&	107				\\
		Avg. radius ($R_{\odot}$)			& 22.83		& 0.04				&	1.14				\\
		Final reduced $\chi^2$			& 2.62												\\
		\hline
	\end{tabular}
\end{table}

\begin{table}
	\caption{Weighting of the different datasets and individual reduced $\chi^2$.}
	\label{table:6}
	\centering
	\renewcommand{\arraystretch}{1.2}
	\begin{tabular}{*{5}{c}}
		\hline
		Observable 			& Ref.		& $\rm{N}_{\rm{points}}$ & NF	& $\chi^2$\\
		\hline
		\hline
		$V_{Johnson}$		& (a)			& 162	& 1.546	& 1.42 \\
		$V_{Walraven}$		& (b)			& 16		& 0.486	& 1.43 \\
		$V_{Tycho}$			& (c, d)		& 163	& 1.550	& 4.37 \\
		$B_{Tycho}$			& (c, d)		& 163	& 1.550	& 6.45 \\
		$HP_{B2000}$			& (c, d)		& 182	& 4.914*	& 6.97 \\
		$J_{CTIO}$			& (e)			& 25		& 0.607	& 1.46 \\
		$H_{CTIO}$			& (e)			& 25		& 0.607	& 0.51 \\
		$K_{CTIO}$			& (e)			& 25		& 0.607	& 1.94 \\
		\hline
		$V_{Walraven}-B_{Walraven}$	& (b)	& 16		& 0.486	& 8.08 \\
		$B_{Johnson}-V_{Johnson}$		& (a)	& 56		& 0.909	& 1.51 \\
		$V_{Johnson}-I_{Bessel}$		& (a)	& 164	& 1.555	& 0.57 \\
		$V_{Johnson}-R_{Bessel}$		& (a)	& 27		& 0.631	& 0.95 \\
		\hline
		 					& $(f)^1$		& 4		& 0.243	&  	   \\
		$\rm{v}_{rad}$			& $(f)^2	$	& 7		& 0.321	& 1.43 \\
		 					& (g)			& 38		& 0.749	&  	   \\
		\hline
		$\theta$ (UD)			& (f)			& 12		& 0.421   & 2.30   \\
		\hline
	\end{tabular}
	\tablefoot{References : 
	(a) \cite{2008yCat.2285....0B}; 
	(b) \cite{1964BAN....17..520W}; 
	(c) \citet{1997ESASP1200.....E}; 
	(d) \cite{1997A&A...323L..61V}; 
	(e) \cite{2008MNRAS.386.2115F}; 
	(f) Present work ($^1$CORALIE and $^2$HARPS); 
	(g) \cite{1964BAN....17..520W}.\\
	*The normalisation factor of the \emph{Hipparcos} data has been multiplied by three to balance the very low uncertainties of the measurements, which would otherwise give to this dataset too much weight in the fitting process.}
\end{table}

\section{Discussion\label{Discussion}}


The $p$-factor value of $\kappa$~Pav that we obtain ($p = \pfactor \pm \pfactorerrtot$) is significantly higher than the value of $0.93 \pm 0.11$ proposed by F08. A $p$-factor smaller than unity would imply that the limb darkening of the star is extremely high, and would generally not have a very clear physical explanation. The present value, however, agrees well with recently published Period-$p$-factor relationships, which give for $\kappa$~Pav ($P=9.09$\,d) \emph{p}-factors of $1.27$ \citep{2013A&A...550A..70G}, $1.29 \pm 0.06$ \citep{2012A&A...543A..55N}, and $1.23 \pm 0.10$ \citep{2009A&A...502..951N}.
It is also consistent with the empirical measurements obtained by \citet{2013MNRAS.436..953P} on the LMC Cepheid OGLE-LMC-CEP-0227 ($P=3.90$\,d, $p=1.21 \pm 0.05$), and by \citet{2005A&A...438L...9M} on $\delta$\,Cep ($P=5.37$\,d, $p=1.27 \pm 0.06$).
Our $p$-factor, however, differs from the $p = 1.37$ and $p=1.359 \pm 0.003$ values predicted by \citet{2012Ap&SS.341..115S} and \citet{2012A&A...541A.134N} respectively. 
The average angular diameter is in good agreement with the value derived by \cite{2012A&A...538A..24G} ($\theta_{\rm{UD}} =1.04 \pm 0.04$ mas at $\phi=0.9$). Converted into linear radius for a distance of $d=179$\,pc, we obtain an average photospheric radius of $22.8$\,R$_{\odot}$. \cite{1997AJ....113.1833B} suggest a comparable value of $19 \pm 5$\,R$_{\odot}$, also derived using the parallax of pulsation method.
The average {\tt SPIPS} effective temperature of $T_\mathrm{eff} = 5739$\,K is slightly higher than the typical values found in the literature. In particular, \citet{1989ApJ...342..476L} find 5500\,K, and \citet{2012A&A...538A..24G} find $T_\mathrm{eff} = 5750$\,K at $\phi=0.94$, which corresponds to $\sim 6336$\,K at the same phase in the present study.  
We find an extinction comparable to the value suggested by F08 ($E(B-V)=0.017$). It is important to stress that the relevance of this parameter relies on the choice of a reddening law, which is in the present case the reddening law from \citet{1999PASP11163F}, taken for $R_{\rm{v}}=3.1$ and differs from the methods used in F08. However, the systematic errors of both measurements dominate these low extinction values.


It was suggested that $\kappa$~Pav belongs to a binary system \citep{2008MNRAS.386.2115F}. The contribution of a stellar companion could have a non-negligible influence on the photometry and radial velocity of the star. We checked our interferometric dataset for the possible presence of a secondary component, by fitting a binary star model that takes the PIONIER closure phases into account. This fitting technique is very sensitive to the presence of companions, with a contrast exceeding 100:1 or more \citep{2011A&A...535A..68A, 2013A&A...552A..21G, 2014A&A...561L...3G}, but we did not identify any companion of $\kappa$~Pav. Our data allowed us to obtain an upper limit of 1\% at 5$\sigma$ on the flux ratio between the two components. This result has been derived from the longest observing sequence (about 3 hours of science and calibrators alternations), which allows reaching a good sensitivity and better uv coverage. Considering the low flux ratio limit, it is unlikely that the data used in the present study could have been biased by the presence of a companion.


The \texttt{SPIPS} code also allows us to consider a possible infrared excess in the fitting process, to track the possible presence of a circumstellar envelope. In the case of $\kappa$~Pav we find an excess of $4.5 \pm 0.5$ \% in the $K$ band. However, it does not improve the quality of the fitting process significantly, so we prefer not to conclude anything about the presence of an actual excess. We nevertheless underline that a circumstellar envelope has been found by \citet{2012A&A...538A..24G}, who identified an infrared excess of about 20\% between $10\,\mu$m and $20\,\mu$m. 

\section{Conclusion}

We report the first observational measurement of the projection factor of the type-II Cepheid $\kappa$~Pav. We combined the HST/FGS parallax from \citet{2011AJ....142..187B} with new interferometric observations from the VLTI/PIONIER instrument, and an extensive set of radial velocities and photometry. Because the period of the star shows unpredictable variations on relatively short timescales, a careful phase adjustment was required to phase the different observing epochs properly.
We obtained a value of $\emph{p} = \pfactor \pm \pfactorerrtot$, which agrees with the Period-$p$-factor relation proposed by \citet{2009A&A...502..951N} and with the empirical measurement obtained by \citet{2013MNRAS.436..953P} in the LMC. It is also consistent with the $p$-factor measured by \citet{2005A&A...438L...9M} for $\delta$~Cep. Although the range of periods presently covered by observational $p$-factor measurements is still limited, our result points at a relatively weak dependence of the $p$-factor on the period, because short- and intermediate-period type-II and classical Cepheids likely share the same $p$-factor within $\approx 5\%$.
Observational measurements of the $p$-factor are difficult, but also essential for the calibration of the distance scale. This factor is presently one of the most important fundamental limitations on the accuracy of the parallax of pulsation distances used to calibrate period-luminosity relationships of Cepheids. Radial velocity modulations (though not evident in the present case) can also lead to systematic errors, as discovered recently by \citet{2014A&A...566L..10A}.
Observational estimates of this parameter are also essential for constraining the $p$-factor models.
The \emph{Gaia} satellite is currently measuring accurate parallaxes for a large number of Galactic Cepheids. This will enable a thorough study of the dependence of the $p$-factor with period and other stellar parameters, and provide us with a solid, unbiased calibration of the parallax of pulsation technique.

\begin{acknowledgements}
The authors would like to thank Dr. George Wallerstein for his useful comments. We acknowledge financial support from the ``Programme National de Physique Stellaire" (PNPS) of CNRS/INSU, France.
PK and AG acknowledge support of the French-Chilean exchange programme ECOS-Sud/CONICYT.
AG acknowledges support from FONDECYT grant 3130361. LSz acknowledges support from the ESTEC Contract No.\,4000106398/NL/KML.
This research received the support of PHASE, the high angular resolution partnership between ONERA, the Observatoire de Paris, CNRS, and University Denis Diderot Paris 7. 
This research made use of the Jean-Marie Mariotti Center \texttt{LITpro} service co-developed by CRAL, LAOG, and FIZEAU, and the Jean-Marie Mariotti Center \texttt{Aspro} service\footnote{Available at http://www.jmmc.fr/aspro}.
We used the SIMBAD and VIZIER databases at the CDS, Strasbourg (France), and NASA's Astrophysics Data System Bibliographic Services.
PIONIER is funded by the Universit\'e Joseph Fourier (UJF), the Agence Nationale pour la Recherche (ANR-06-BLAN-0421 and ANR-10-BLAN-0505), and the Institut National des Science de l'Univers (INSU PNP and PNPS). Its beam combiner is from IPAG and CEA-LETI based on CNES R\&T funding.
\end{acknowledgements}


\bibliographystyle{aa} 
\bibliography{kappa_pav}

\end{document}